\documentclass[aps,showpacs,showkeys,preprintnumbers,twocolumn,superscriptaddress]{revtex4}
\usepackage{colordvi}
\usepackage{graphicx}
\usepackage{dcolumn}
\usepackage{amsmath}

\newcommand{\eV}{\mbox{\rm eV}}
\newcommand{\keV}{\mbox{\rm keV}}
\newcommand{\GeV}{\mbox{\rm GeV}}

\begin{document}

\title{Search for hadronic axions using axioelectric effect}
 
\author{A. Ljubi\v{c}i\'{c}}

\author{D.Kekez}
 
\author{Z. Kre\v{c}ak}

\affiliation{Rudjer Bo\v{s}kovi\'{c} Institute,
   Bijeni\v{c}ka 54, 10001 Zagreb, Croatia}
 
\author{T. Ljubi\v{c}i\'{c}}
 
\affiliation{Physics Department, Brookhaven National Laboratory, Upton, N.Y., USA}

\begin{abstract}
We made a search for hadronic axions which could be emitted 
from the Sun in M1 transitions between the first 14.4 keV thermally 
excited and the ground state in $^{57}$Fe, and absorbed in the HPGe 
detector by axioelectric effect. An upper limit on hadronic axion 
mass of 400 eV is obtained at the 95\% confidence level.
\end{abstract}

\pacs{14.80.Mz}
\keywords{hadronic axions search, solar axions, axioelectric effect}

\maketitle
 
 Axions arise as a result of spontaneous breaking of the Peccei--Quinn 
(PQ) chiral symmetry \cite{b1}. This symmetry was introduced to resolve 
the strong CP problem associated with the $\Theta$--vacuum structure of QCD.
The solution predicts the existence 
of a neutral spin--zero pseudoscalar particle, called axion, with 
a non--zero mass $m_a$. The axion mass
can be interpreted as a mixing of the axion field with pions 
and is related to the PQ symmetry breaking scale $f_a$
by $m_a f_a \approx m_\pi f_\pi$ where $m_\pi=135$ MeV is the pion mass and 
$f_\pi=93$ MeV its decay constant. The axion mass, or the PQ symmetry 
breaking scale, is arbitrary and all values solve the strong 
CP problem. The original PQ suggestion that $f_a$
is equal to the electroweak symmetry breaking scale of 250 GeV, 
which requires an axion mass $m_a$
of few hundred keV was quickly ruled out by experiment. The 
constrains on the axion mass from various laboratory experiments, 
astrophysics and cosmology restrict the masses of axion to a 
narrow $10^{-5}\eV\le m_a \le 10^{-2}\eV$
range. New axion models are introduced with breaking scale $f_a$
at values of $10^{10}$--$10^{12}$ GeV which escapes all the phenomenological 
constrains. Therefore, all axion couplings become extremely small 
and axion models of this type are referred to as invisible axion 
models. Two classes of invisible axion models have been developed: 
KSVZ (Kim, Shifman, Vainshtein and Zakharov) or hadronic axion 
\cite{b2} and DFSZ (Dine, Fischler, Srednicki and Zhitnitsky) or GUT 
axion \cite{b3} model.


The main difference between KSVZ and DFSZ axions is that the 
former have no tree--level couplings to ordinary quarks and leptons 
because new fermions have been introduced that carry the PQ charge 
while usual quarks and leptons do not. As a result, the interaction 
of KSVZ--type axions with electrons is strongly suppressed. However, 
their coupling to nucleons is not zero due to the axion--pion 
mixing which exists even if the tree--level coupling to ordinary 
quark vanishes. Axions are also one of the best known and most 
studied candidates for the cold dark matter in the Universe. 
Axions with mass $\approx 10^{-5}~\eV$
would make a significant component of cold dark matter.


Besides the $10^{10}$--$10^{12}$ GeV there is another window around $10^{6}$ 
GeV. This is the window for hadronic axions only. For this window 
the existing constraints based on the axion--photon coupling are 
excluded. In Kim's composite axion models \cite{b4} the axion--photon 
coupling is model dependent and may be significantly reduced. 
It was shown by Kaplan \cite{kaplan85} that as a result the astrophysical 
bounds on the axion mass are weakened and certain Kim's models 
allow axion mass as large as $\approx 20~\eV$. 
Other astrophysical and cosmological bounds place a lower limit 
on the hadronic axion mass $m_a\ge 10~\eV$ \cite{b6}.
Recently, axions in this narrow hadronic axion mass window 
of $10~\eV\le m_a\le 20~\eV$ 
have been proposed as candidates for hot dark matter of the 
Universe \cite{b7}.


A possible source of axions is our Sun. Solar axion spectra would 
consist of the continuous part, generated through the Primakoff 
effect and line spectra generated mostly in nuclear M1 transitions 
of some nuclides. Huxton and Lee \cite{b8} have calculated axion emission 
rates from the Sun for the M1 transitions in $^{57}$Fe, $^{55}$Mn and 
$^{23}$Na. Other convenient sources of monoenergetic solar axions 
are M1 transitions in $^{7}$Li \cite{b9} and 
$^{83}$Kr \cite{b10}. Estimates suggest that the highest emission rate of 
monoenergetic axions will be produced during the 14.4 keV M1 
transition in $^{57}$Fe. Moriyama \cite{b11} proposed to detect these 
axions using the resonant absorption process by the same nuclide 
in the Earth--bound laboratory. Since both emission and absorption 
proceed via axion--nucleon coupling this method is free from the 
uncertainty of the axion--photon coupling. Using this approach 
Kr\v{c}mar {\em et al}. \cite{b12} obtained an upper limit of 
$m_a\le 745~\eV$.
We have also searched for 477.6 keV and 9.4 keV axions, which 
were supposed to be emitted from the Sun during M1 transitions 
in $^{7}$Li and $^{83}$Kr and resonantly absorbed by the same nuclides 
in our laboratory. We did not observe such events and only upper 
limits on the hadronic axion mass of 32 keV \cite{b9} and 5.5 keV \cite{b10} 
at the 95\% confidence level were obtained. 


 Other possible interactions for detecting hadronic axions are: 
(i) Compton conversion of an axion to a photon $a+e\to\gamma+e$,
and (ii) axioelectric effect $a+e+Z\to e+Z$.
Former process was first considered by Donnelly {\em et al}. \cite{b13} 
and the latter one by Zhitnitsky and Skovpen \cite{b14}. Even if there 
is no axion--electron coupling at the tree--level for the KSVZ 
models, there will be an induced axion--electron coupling at the 
one--loop level, which was calculated by Srednicki \cite{srednicki85},

\begin{equation}
g_{aee}
=
2.2\times 10^{-15} m_a(\eV) N
\left[
\bar{c}_{a\gamma\gamma}\ln\frac{f_a}{m}
-
\frac{2}{3}\frac{4+Z}{1+Z}\ln\frac{\Lambda}{m}
\right]~.
\label{g_aee}
\end{equation}

\noindent Here $N$ is number of generations, $Z=m_u/m_d\approx 0.55$
is the $u$ and $d$ quark mass ratio, $\Lambda\approx 1~\GeV$
is the cutoff of order of the chiral symmetry breaking scale, 
$\bar{c}_{a\gamma\gamma}$ is ratio of electromagnetic and color anomalies,
$f_a=f_\pi m_\pi \sqrt{Z}/(m_a (1+Z))$,
and m is electron mass.
For the hadronic axions with mass range 
$10~\eV\le m_a\le 20~\eV$ Kaplan \cite{kaplan85} has shown that 
$\bar{c}_{a\gamma\gamma}=2$. In his minimal composite axion model \cite{b4} Kim
considered the case with $N=3$. Introducing these values into (\ref{g_aee})
we obtain
$g_{aee} = 6.6\times 10^{-15} m_a(\eV)
[-14.83+2\ln(1.176\times 10^{10}/m_a(\eV))]$.


Coupling of hadronic axions with electrons is much weaker than 
with nucleons. However, the high temperature in the Sun's interior 
will broaden the nuclear line width in the Sun and, in the experiments 
based on the resonant absorption, this reduces the effective 
axion flux by a factor 
$\Gamma_{\mbox{\rm\scriptsize Earth}}/\Gamma_{\mbox{\rm\scriptsize Sun}}\ll 1$;
$\Gamma_{\mbox{\rm\scriptsize Sun}}$ and 
$\Gamma_{\mbox{\rm\scriptsize Earth}}$
are the decay widths of the same nuclear state in the Sun and 
in the Earth--bound laboratory. In the axion--to--photon Compton 
conversion and the axioelectric effect all axions emitted during 
nuclear transition in the Sun will contribute, and this will 
compensate the smallness of the axion--electron coupling constant. 


Using theoretical predictions of Zhitnitsky and Skovpen \cite{b12} 
we estimate that for the 14.4 keV axions the cross section for 
the axioelectric effect is about three orders of magnitude larger 
than the cross section for the axion--to--photon Compton conversion 
process. Therefore we will further consider only the axioelectric 
effect. In this effect an axion dissappears and an electron is 
ejected from an atom. The electron carries away all the energy 
of the absorbed photon, minus the energy binding the electron 
to the atom. Zhitnitsky and Skovpen \cite{b12} calculated cross sections 
for the axioeffect for atoms with $Z\ll 137$,
for small axion masses and for the axion energy large compared 
to the K--electron binding energy. For the cross section on the 
K--electron they obtained 

\begin{eqnarray}
\sigma_{ae\to e}
=
\frac{8\pi\alpha_{ae}}{m^2}
(Z\alpha m)^5
\frac{p_e}{k_a}
\left[
\frac{4\omega(\omega^2+m_a^2)}{(k_a^2-p_e^2)^4}
\right. \nonumber \\ \left.
-\frac{2\omega}{(k_a^2-p_e^2)^3}
-\frac{64}{3}p_e^2 k_a^2 m \frac{m_a^2}{(k_a^2-p_e^2)^6}
-\frac{16 m_a^2 k_a^2 \epsilon}{(k_a^2-p_e^2)^5}
\right. \nonumber \\ \left.
-\frac{\omega}{p_e k_a}\frac{1}{(k_a^2-p_e^2)^2}
   \ln\left(\frac{p_e+k_a}{p_e-k_a}\right)
\right]~,
\end{eqnarray}

\noindent where $\alpha=e^2/(4\pi)=1/137$, $\alpha_{ae}=g_{aee}^2/(4\pi)$;
$\omega$, $k_a$ and $\epsilon$, $p_e$ are the energy and the absolute value
of the momentum of the axion and the ejected electron, respectively.

   In the case of $m_a\to 0$ and for axion energies 
$\epsilon\ll m$ the cross section becomes 

\begin{equation}
\sigma_{ae\to e}
=
\sqrt{2}
\left(\frac{\alpha_{ae}}{\alpha}\right)
\left(\frac{8\pi}{3} \frac{\alpha^2}{m^2}\right)
\alpha^4 Z^5 \left(\frac{m}{\omega}\right)^{\frac{3}{2}}~.
\end{equation}


 In our investigations of solar hadronic axions we have searched 
for a peak at 14.4 keV in a single spectrum recorded in a HPGe 
detector. If observed, this peak could be interpreted as the 
result of the axioelectric effect of the 14.4 keV axions on germanium 
atoms, with subsequent absorption of the emitted electrons
and accompanying x rays in the crystal. The 14.4 keV axions are 
supposed to be emitted from the Sun in M1 transition between 
the first thermally excited state and the ground state in $^{57}$Fe. 
In this experimental set--up the target and the detector are the 
same and the efficiency of the system is substantially increased. 
$^{57}$Fe is one of the stable isotopes of iron (natural abundance 
2.2\%) which is exceptionally abundant among heavy elements in 
the Sun (solar abundance by mass fraction $2.7\times 10^{-5}$ ).
Also the 14.4 keV level is relatively easy to excite 
thermally because its energy is comparable with the temperature 
in the Sun core ($\approx 1.3~\keV$).
The axion flux from the Sun was estimated by Moriyama 
\cite{b11} using similar calculation as in Ref. \cite{b8}. He obtained the 
value of 
$d\Phi(\epsilon)/d\epsilon
=
1.7\times 10^{10} \times m_a^2(\eV)
{\mbox{\rm cm}}^{-2} {\mbox{\rm s}}^{-1}\keV^{-1}$.
As a result of the Doppler broadening in the Sun the estimated 
width of the 14.4 keV peak is 
$\Gamma_{\mbox{\rm\scriptsize Sun}}\approx 5~\eV$.
Therefore the total flux of 14.4 keV axions is expected to 
be $\Phi=8.5\times 10^7 m_a^2(\eV) {\mbox{\rm cm}}^{-2} {\mbox{\rm s}}^{-1}$. 


 In the search for the 14.4 keV peak we have re--analyzed single 
spectra accumulated in the HPGe detector during our investigation 
of solar hadronic axions using $^{7}$Li. These axions were suggested 
to be created in M1 transitions between the first excited 478 
keV and the ground state in $^{7}$Li and were expected to excite 
resonantly the same energy level in the $^{7}$Li target, placed 
in the laboratory. The subsequently emitted gamma rays were searched 
for in a HPGe detector. Run spectra were obtained with the 56.72 
g lithium target placed in front of the detector while for the 
background spectra the lithium target was replaced with an appropriate 
absorber simulating background conditions. Run and background 
were both counted during the collection time $t=111.11$
days each. Attenuation of axions in detector shieldings is negligible 
and we could sum both run and background spectra and improve 
the statistics by duplicating the data collection time. HPGe 
detector was placed in a 13 cm deep and 8 cm in diameter well 
of a 25.4 cm \ensuremath{\times} 20.3 cm in diameter NaI crystal. The crystal 
was located inside an iron box with internal dimensions
$54\times 33 \times 33$ cm$^{3}$
and with wall thickness ranging from 16 to 23 cm. 
The iron was more than 65 years old and was essentially free of $^{60}$Co 
impurities. The box was lined outside with 1 cm thick lead. HPGe 
detector operated in anticoincidence mode with NaI crystal and 
the background events around 14.4 keV were reduced by a factor 
of $\approx 10$.

\begin{figure}[b]
\includegraphics[width=200mm,angle=0]{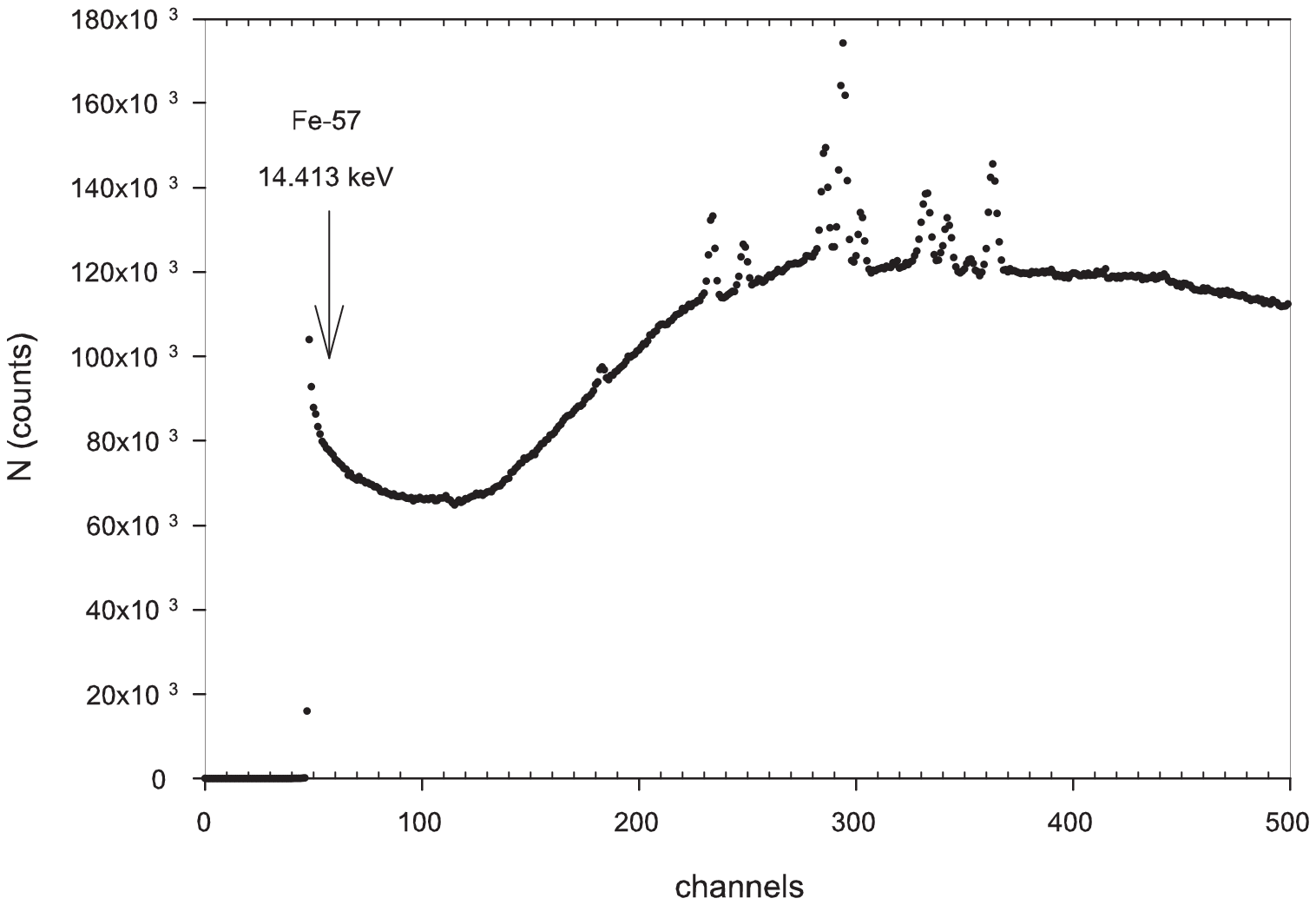}
\caption{}
\label{Fig1}
\end{figure}

The energy spectra accumulated for the 222.22 days are shown 
in Fig.\ref{Fig1}. We have obtained an energy calibration with a set of 
calibrated radioactive sources. The 14.4 keV peak is expected 
at the 55$^{\mbox{\rm\scriptsize th}}$ channel.
Energy resolution (FWHM) at this energy 
was estimated by extrapolating energy resolutions obtained from 
the higher energy region. The FWHM at 14.4 keV was estimated 
to be 0.9 keV which corresponds to 3.5 channels.

We can relate the axion mass with the number $N_a$
of counts, detected at 14.4 keV in our HPGe crystal, with the 
expression

\begin{equation}
N_a
=
\Phi \sigma_{ae\to e}\,2\, N_{\mbox{\rm\scriptsize Ge}}\, \varepsilon_{14.4}\, t~.
\end{equation}

\noindent Number of germanium atoms in our 50 mm in diameter $\times$
40 mm thick
HPGe crystal is $N_{\mbox{\rm\scriptsize Ge}}=3.47\times 10^{24}$.
Factor 2 is number of electrons in K--shell.
The intrinsic efficiency $\varepsilon_{14.4}$
of our detector for the electrons and accompanying germanium 
x rays is $\approx 1$. The data were collected during $t=1.92\times 10^7$ s.
Introducing calculated $g_{aee}$ into (2) we obtain for the cross section
for the axioelectric effect on germanium atom 
$\sigma_{ae\to e}=1.80\times 10^{-50}\, m_a^2(\eV)
{}[-14.83+2\ln(1.176\times 10^{10}/ m_a(\eV))]^2 {\mbox{\rm cm}}^2$.
Then from Eq. (4) we obtain for the expected number of the axioeffect 
events $N_a=1.02\times 10^{-10} m_a^4(\eV)
{}[-14.83+2\ln(1.176\times 10^{10}/ m_a(\eV))]^2$.

  We could not identify a peak at 14.4 keV. For the number of axioeffect 
events we used the number of counts detected in 6 channels at 
the 14.4 keV region and obtained an upper limit of $N_a \le 980$
events. This gives an upper limit on the hadronic axion mass 
of $m_a\le 400~\eV$ at the 95\% confidence level.

  F. T. Avignone III {\em et al.} \cite{avignone98} searched for a signal
of solar axions coherently converting into photons via the Primakoff effect.
Their analysis yield a laboratory bound to $a\gamma\gamma$ coupling
of $g_{a\gamma\gamma}<2.7\times 10^{-9} \GeV^{-1}$.
The $a\gamma\gamma$ coupling is model dependent and it is given by
 
\begin{eqnarray}
g_{a\gamma\gamma}
=
\frac{\alpha}{\pi f_a}
\frac{1}{2}\left[\bar{c}_{a\gamma\gamma}-\frac{2(4+Z)}{3(1+Z)} \right]
\nonumber \\
=
\frac{\alpha}{\pi f_a}
\frac{1}{2}\left[\bar{c}_{a\gamma\gamma}-1.954\pm 0.036 \right]~,
\label{Gagg}
\end{eqnarray}

\noindent where the value of the second term in Eq.~(\ref{Gagg}) and its
error is calculated using $Z=0.553\pm 0.043$ \cite{leutwyler96b}.
For KSVZ models it is easy to construct models in which
$\bar{c}_{a\gamma\gamma}=2$
\cite{kaplan85}.
This leads to great suppression of axion--photon coupling $g_{a\gamma\gamma}$
because of a cancellation between two terms in Eq.~(\ref{Gagg}) \cite{kaplan85}.
Moreover, Particle Data Group \cite{PDG02} quotes $0.3 < Z < 0.7$ allowing even
vanishing axion--photon coupling.
   In KSVZ models the axion--electron coupling is radiatively induced at the
one--loop level \cite{srednicki85}. The corresponding diagrams are
logarithmically divergent. The different cutoff scales of the part
proportional to $E/N$ and the part proportional to ${2(4+Z)}/[3(1+Z)]$
result in an axion--electron coupling $g_{aee}$,
Eq.~(\ref{g_aee}), which remains finite even in the case when
$g_{a\gamma\gamma}$ is very small (or possibly vanish).
In this particular case our measurement, based on axion--electron interaction,
gives an upper limit to the hadronic axion mass which is rather
insensitive to the value of Z, while the result of Ref.~\cite{avignone98}
cannot give reliable prediction about hadronic axion mass.

 Because the number of detected gamma rays is proportional to 
the fourth power of axion mass it is very difficult to improve 
the sensitivity of the detection system. One should have a detector 
of much larger volume and, at the same time, the background in 
the 14.4 keV region should be reduced by a very large factor. An 
obvious choice would be a large volume gaseous Time Projection 
Chamber (TPC). One such detector is the STAR experiment's TPC 
currently operating at the Relativistic Heavy Ion Collider (RHIC) 
in Upton, USA. This TPC is 4 m long \ensuremath{\times} 4 m in diameter 
\cite{b16} and is filled with an Ar + CH$_{4}$ mixture operating under 
atmospheric pressure. The active drift time of this device is 
\ensuremath{\sim}50 microseconds while the readout time is presently 
10 milliseconds. This ratio reduces the overall efficiency of 
this system by about 200. In an axioelectric effect with the 
14.4 keV axion, an electron with \ensuremath{\sim} 11 keV will be emitted 
together with argon x rays. The electron will be stopped within 
\ensuremath{\sim} 0.3 cm of the gas and will thus give a clear, pointlike 
signal in the volume of the TPC. The energies of the argon x rays 
are too low and will escape undetected. The pointlike nature 
of the electron signal as well as its narrow energy is expected 
to be easily distinguished from other background sources ({\em i.e.} 
cosmic rays and other higher momentum particles) which leave 
an easily recognizable extended track segment. The STAR TPC is 
surrounded by standard scintillator strips along the whole outer 
surface of the barrel as well as scintillator panels covering 
a part of the detector's base. These auxiliary detectors could 
be used in anti--coincidence ({\em i.e.} as a veto against charged tracks 
traversing the volume of the TPC) during STAR's normal data taking 
periods (almost 6 months/year) thus enabling a parasitic axion 
search while the STAR TPC is actively involved in RHIC's heavy--ion 
program. The sensitivity on the axion mass of \ensuremath{\sim} 210 eV 
could be reached under reasonable assumptions of \ensuremath{\sim} 1 
background count/day operating for 1 year with the STAR TPC.


In the near future the STAR collaboration is expected to improve 
the data--acquisition readout time of the TPC by a factor \texttt{<} 
50 which would then considerably increase the efficiency of the 
proposed measurement.
Because the axioeffect cross section is proportional to Z$^{5}$ the 
efficiency of the system could be further improved by filling 
the TPC with the krypton gas. In that case if the TPC operates 
for one year with the background rate of 
$\approx 1$ count/day, we would obtain an upper limit on $m_a$
of $\approx 10~\eV$, which would close this hadronic axion mass window.

\section*{Acknowledgments}

The authors wish to thank the Ministry of Science and Technology 
of Croatia for financial support. One of the authors (T.L.) is 
grateful for the support of Brookhaven National Laboratory.

\end{document}